\documentclass[pra,amssymb,showpacs,showkeys,twocolumn]{revtex4}
\usepackage[T1]{fontenc}
\usepackage{textcomp}
\usepackage{graphicx}
\usepackage{longtable}
\RequirePackage{times}
\RequirePackage{mathptm}
\usepackage[usenames]{color}
\newcommand{\Red}{\color{Red}}  
\newcommand{\Green}{\color{Green}}  
\def\ddarrow{{\swarrow\mkern-18mu\nearrow}}
\def\cddarrow{{\searrow\mkern-18mu\nwarrow}}
\begin{document}

\title{Staging quantum cryptography with chocolate balls\footnote
{The author reserves the copyright for all public performances.
Performance lincenses are granted for educational institutions
and other not-for-profit performances for free;
these institutions are kindly asked to send a small note about the performance
to the author.
}}

\author{Karl Svozil}
\email{svozil@tuwien.ac.at}
\homepage{http://tph.tuwien.ac.at/~svozil}
\affiliation{Institut f\"ur Theoretische Physik, University of Technology Vienna,
Wiedner Hauptstra\ss e 8-10/136, A-1040 Vienna, Austria}

\begin{abstract}
Moderated by a director, laymen and students are encouraged to assume the role of quanta and enact a quantum cryptographic protocol.
The performance is based on a generalized urn model capable of reproducing complementarity even for classical chocolate balls.
\end{abstract}


\maketitle

\begin{quote}
\begin{flushright}
{
Dedicated to Antonin Artaud, \\
author of {\it Le th{\'{e}}{\^{a}}tre et son double} \cite{Arthaud}.
 }
\end{flushright}
\end{quote}

\section{Background}

Quantum cryptography is a relatively recent and extremely active field of
research within quantum physics. Its main characteristic is the use of (at least ideally)
individual particles for encrypted information transmission. Its objective
is to encrypt messages, or to create and enlarge a set of secret equal
random numbers, between two spatially separated agents by means of elementary
particles, such as single photons, which are transmitted through a quantum
channel.

The history of quantum cryptography dates back to around 1970, to the manuscript
by Wiesner \cite{wiesner} and a  protocol
by Bennett {\&} Brassard in 1998 \cite{benn-82,benn-84,ekert91,benn-92,gisin-qc-rmp}
henceforth called ``BB84''.
Since
then, experimental prototyping has advanced rapidly.
Without going into too
much detail and just to name a few examples, the work ranges from the very
first experiments carried out in the IBM Yorktown Heights Laboratory
by Bennett and co-workers  in
1989 \cite{benn-92},
to signal transmissions across Lake Geneva in 1993 \cite{gisin-qc-rmp},
and the
network in the Boston Metropolitan Area which has been sponsored by DARPA
since 2003 \cite{ell-co-05}.
In a much publicized, spectacular demonstration,
a quantum cryptographic aided bank transfer took place via optical fibers
installed in the sewers of Vienna in the
presence of some local politicians and bank representatives \cite{pflmubpskwjz}.

Quantum cryptography forms an important link
between quantum theory and experimental technology, and
possibly even industrial applications.
The public is highly interested in quantum physics and quantum
cryptography, but the protocols used are rarely made available to the layman or student
in any detail. For an outsider these subjects seem to be shrouded in some
kind of "mystic veil" and are very difficult to understand,
although great interest in the subject prevails.

In what follows, we shall use a simple but effective generalized urn model introduced by Wright
\cite{wright,wright:pent,svozil-2001-eua} to mimic complementarity.
A generalized urn model
is
characterized by an ensemble of balls with black background color.
Printed on these balls are some color symbols from a symbolic alphabet.
The colors are elements of a set of colors.
A particular ball type is associated with a unique combination of mono-spectrally
(no mixture of wavelength) colored symbols
printed on the black ball background.
Every ball contains just one single symbol per color.

Assume further some mono-spectral filters or eyeglasses which are
``perfect'' by totally absorbing light of all other colors
but a particular single one.
In that way, every color can be associated with a particular eyeglass and vice versa.

When a spectator looks at a particular ball through such an eyeglass,
the only operationally recognizable symbol will be the one in the particular
color which is transmitted through the eyeglass.
All other colors are absorbed, and the symbols printed in them will appear black
and therefore cannot be differentiated from the black background.
Hence the ball appears to carry a different ``message'' or symbol,
depending on the color at which it is viewed.
We will present an explicit example featuring complementarity, in very similar ways as
quantum complementarity.

The difference between the chocolate balls and the quanta is the possibility
to view all the different symbols on the chocolate balls
in all different colors by taking off the eyeglasses.
Quantum mechanics does not provide us with such a possibility.
On the contrary, there are strong formal arguments suggesting
that the assumption of a simultaneous
physical existence \cite{epr} of such complementary observables yields a complete contradiction
\cite{kochen1}.

\section{Principles of conduct}

In order to make it a
real-life experience, we have aimed at dramatizing quantum cryptography.
The quantum world is turned into a kind of drama, in which actors and
a moderator present a quantum cryptographic protocol on stage. The audience
is actively involved and invited to participate in the dramatic
presentation. If at all possible, the event should be moderated by a
well-known comedian, or by a physics teacher.

The entire process is principally analogous to an
experiment in a slightly surreal sense: just like humans, single quanta
are never completely predictable. Among other things,
they are in fact determined by random events, and marked by a certain
``noise'' similar to the chaos that will certainly accompany the public
presentation of the quantum cryptographic protocols. Therefore, the
interference of individual participants is even encouraged and not
a deficiency of the model.

Throughout the performance, everybody should have fun, relax, and try
to feel and act like an elementary particle -- rather in the spirit of the
meditative Zen koan ``Mu.'' The participants might manage to feel
like Schr\"{o}dinger's cat \cite{schrodinger},
or like a particle simultaneously passing
through two spatially separated slits.
In idle times, one may even contemplate how conscious minds could experience a coherent
quantum superposition between two states of consciousness.
However, this kind of sophistication is neither necessary, nor particularly important for
dramatizing quantum cryptographic protocols.

Our entire empirical knowledge of the world is based on the occurrence
of elementary (binary) events, such as the reactions caused by quanta in
particle detectors yielding either a ``click'' or none.
Therefore, the following simple syntactic rules should not be dismissed as
mere cooking recipes, for quantum mechanics itself can actually be
applied merely as a
sophisticated set of laws with a possibly superfluous \cite{fuchs-peres}
semantic superstructure.

\section{Instructions for staging the protocol}

Our objective is to generate a secret sequence of random numbers only known
by two agents called Alice and Bob. In order to do so, the following
utensils depicted in Figure \ref{2005-ln1e-utensils} will be required:
\begin{figure}
\begin{center}
 \includegraphics[width=8.2cm]{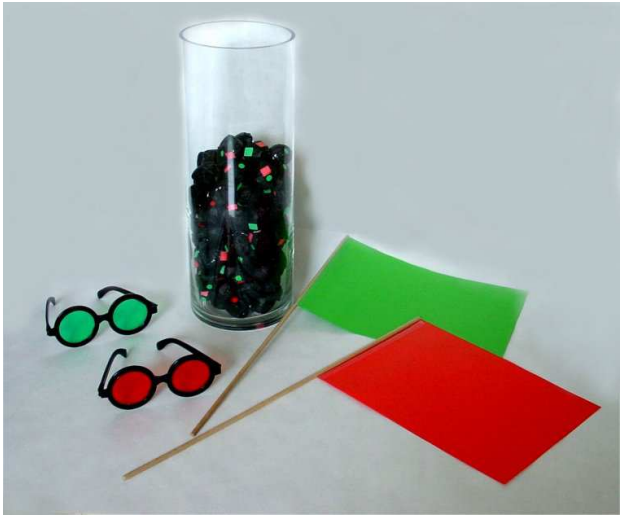}
\end{center}
   \caption{Utensils required for staging the BB84 protocoll.}
\label{2005-ln1e-utensils}
 \end{figure}

\begin{itemize}
\item[(1)]
Two sets each of fully saturated glasses in red and green (complementary
colors)

\item[(2)]
An urn or bucket

\item[(3)]
A large number of foil-wrapped chocolate balls (in Austria called ``Mozartkugeln'')
or similar -- each with a black background,
imprinted with one red and one green symbol (either 0 or 1) -- to be placed
inside the urn. According to all possible combinations, there are four types
altogether, which can be found in Table \ref{2005-nl1-t1}. There needs to be an equal number
of each type in the urn.
\begin{table}
\begin{tabular}{ccc}
\hline\hline
Balltyp&{\Red red}&{\Green green}\\
\hline
Typ 1&{\Red 0}&{\Green 0}\\
Typ 2&{\Red 0}&{\Green 1}\\
Typ 3&{\Red 1}&{\Green 0}\\
Typ 4&{\Red 1}&{\Green 1}\\
\hline\hline
\end{tabular}
\caption{Schema of imprinting of the chocolate balls.\label{2005-nl1-t1}}
\end{table}

\item[(4)]
Small red and green flags, two of each

\item[(5)]
Two blackboards and chalk (or two secret notebooks)

\item[(6)]
Two coins
\end{itemize}

The following acting persons are involved:

\begin{itemize}
\item[(1)]
A moderator who makes comments and ensures that the participants more or
less adhere to the protocol as described below. The moderator has many liberties and
may even choose to stage cryptographic attacks.

\item[(2)]
Alice and Bob, two spatially separated parties

\item[(3)]
Ideally, but not necessary are some actors who know
the protocol and introduce new visitors to the roles of
Alice, Bob and the quanta.

\item[(4)]
A large number of people assuming the roles of the quanta. They are in
charge of transmitting the chocolates and may eat them in the course of events
or afterwards.
\end{itemize}

In our model, chocolates marked with the symbols 0 and 1 in red, correspond
to what in quantum optics correspond to horizontally ($\leftrightarrow$)
and vertically ($\updownarrow$) polarized photons, respectively.
Accordingly, chocolates marked with the symbols 0 and 1 in green, correspond
to left ($\circlearrowleft$) and right ($\circlearrowright$)
circularly polarized photons, or alternatively to linearly
polarized photons with polarization directions ($\ddarrow$) and ($\cddarrow$)
rotated by 45$°$ ($\pi / 4$) from the horizontal and the vertical,
respectively.

The protocol is to be carried out as follows:

\begin{itemize}
\item[(1)]
Alice flips a coin in order to chose one of two pairs of glasses: heads is
for the green glasses, tails for the red ones. She puts them on and randomly
draws one chocolate from the urn. She can only read the symbol in the color
of her glasses (due to subtractive color the other symbol in the
complementary color appears black and cannot be differentiated from the black background).
This situation is illustrated in Figure \ref{f-gum-w}. She
writes the symbol she could read, as well as the color used,
either on the blackboard or into her
notebook. Should she attempt to take off her glasses or look at the symbols
with the other pair, the player in the role of the quantum is required to eat
the chocolate at once.
\begin{figure}
\begin{center}
\unitlength 0.6mm
\linethickness{0.4pt}
\begin{picture}(118.33,55.67)
\put(35.00,35.00){\line(0,-1){25.00}}
\put(55.00,9.67){\oval(40.00,9.33)[b]}
\put(75.00,35.00){\line(0,-1){25.00}}
\put(40.62,10.00){\circle*{9.20}}
\put(40.62,10.00){\makebox(0,0)[cc]{\bf {\Red 1} {\Green 0}}}
\put(50.29,10.00){\circle*{9.20}}
\put(50.29,10.00){\makebox(0,0)[cc]{\bf {\Red 1} {\Green 1}}}
\put(60.33,10.00){\circle*{9.20}}
\put(69.91,10.00){\circle*{9.20}}
\put(60.33,10.00){\makebox(0,0)[cc]{\bf {\Red 0} {\Green 1}}}
\put(69.91,10.00){\makebox(0,0)[cc]{\bf {\Red 0} {\Green 0}}}
\put(46.33,19.00){\circle*{9.20}}
\put(46.33,19.00){\makebox(0,0)[cc]{\bf {\Red 0} {\Green 1}}}
\put(56.00,18.67){\circle*{9.20}}
\put(56.00,18.67){\makebox(0,0)[cc]{\bf {\Red 1} {\Green 1}}}
\put(65.81,19.00){\circle*{9.20}}
\put(65.81,19.00){\makebox(0,0)[cc]{\bf {\Red 1} {\Green 0}}}
\put(40.15,26.33){\circle*{9.20}}
\put(40.15,26.33){\makebox(0,0)[cc]{\bf {\Red 0} {\Green 1}}}
\put(51.76,27.57){\circle*{9.20}}
\put(51.76,27.57){\makebox(0,0)[cc]{\bf {\Red 0} {\Green 0}}}
\put(69.66,27.91){\circle*{9.20}}
\put(69.66,27.91){\makebox(0,0)[cc]{\bf {\Red 1} {\Green 0}}}
\put(5.00,41.00){\Red \circle*{5.13}}
\put(13.67,41.00){\Red \circle*{5.13}}
\put(9.33,42.17){\oval(3.33,2.33)[t]}
\multiput(16.33,41.00)(0.11,0.16){16}{\line(0,1){0.16}}
\multiput(18.16,43.55)(0.12,0.15){14}{\line(0,1){0.15}}
\multiput(19.79,45.67)(0.11,0.13){13}{\line(0,1){0.13}}
\multiput(21.24,47.38)(0.11,0.12){11}{\line(0,1){0.12}}
\multiput(22.49,48.67)(0.13,0.11){8}{\line(1,0){0.13}}
\multiput(23.56,49.54)(0.22,0.11){4}{\line(1,0){0.22}}
\put(24.43,49.99){\line(1,0){0.68}}
\multiput(25.11,50.02)(0.12,-0.10){4}{\line(1,0){0.12}}
\multiput(25.60,49.64)(0.10,-0.27){3}{\line(0,-1){0.27}}
\put(25.90,48.83){\line(0,-1){1.83}}
\multiput(2.33,41.00)(0.11,0.16){16}{\line(0,1){0.16}}
\multiput(4.16,43.55)(0.12,0.15){14}{\line(0,1){0.15}}
\multiput(5.79,45.67)(0.11,0.13){13}{\line(0,1){0.13}}
\multiput(7.24,47.38)(0.11,0.12){11}{\line(0,1){0.12}}
\multiput(8.49,48.67)(0.13,0.11){8}{\line(1,0){0.13}}
\multiput(9.56,49.54)(0.22,0.11){4}{\line(1,0){0.22}}
\put(10.43,49.99){\line(1,0){0.68}}
\multiput(11.11,50.02)(0.12,-0.10){4}{\line(1,0){0.12}}
\multiput(11.60,49.64)(0.10,-0.27){3}{\line(0,-1){0.27}}
\put(11.90,48.83){\line(0,-1){1.83}}
\put(97.00,41.33){\Green \circle*{5.13}}
\put(105.67,41.33){\Green \circle*{5.13}}
\put(101.33,42.50){\oval(3.33,2.33)[t]}
\multiput(108.33,41.33)(0.11,0.16){16}{\line(0,1){0.16}}
\multiput(110.16,43.88)(0.12,0.15){14}{\line(0,1){0.15}}
\multiput(111.79,46.01)(0.11,0.13){13}{\line(0,1){0.13}}
\multiput(113.24,47.72)(0.11,0.12){11}{\line(0,1){0.12}}
\multiput(114.49,49.01)(0.13,0.11){8}{\line(1,0){0.13}}
\multiput(115.56,49.88)(0.22,0.11){4}{\line(1,0){0.22}}
\put(116.43,50.33){\line(1,0){0.68}}
\multiput(117.11,50.36)(0.12,-0.10){4}{\line(1,0){0.12}}
\multiput(117.60,49.97)(0.10,-0.27){3}{\line(0,-1){0.27}}
\put(117.90,49.16){\line(0,-1){1.83}}
\multiput(94.33,41.33)(0.11,0.16){16}{\line(0,1){0.16}}
\multiput(96.16,43.88)(0.12,0.15){14}{\line(0,1){0.15}}
\multiput(97.79,46.01)(0.11,0.13){13}{\line(0,1){0.13}}
\multiput(99.24,47.72)(0.11,0.12){11}{\line(0,1){0.12}}
\multiput(100.49,49.01)(0.13,0.11){8}{\line(1,0){0.13}}
\multiput(101.56,49.88)(0.22,0.11){4}{\line(1,0){0.22}}
\put(102.43,50.33){\line(1,0){0.68}}
\multiput(103.11,50.36)(0.12,-0.10){4}{\line(1,0){0.12}}
\multiput(103.60,49.97)(0.10,-0.27){3}{\line(0,-1){0.27}}
\put(103.90,49.16){\line(0,-1){1.83}}
\put(7.67,23.67){\circle*{9.20}}
\put(7.67,23.67){\makebox(0,0)[cc]{\bf \Red ?}}
\put(101.33,24.33){\circle*{9.20}}
\put(101.33,24.33){\makebox(0,0)[cc]{\bf \Green ?}}
\put(101.33,0.00){\makebox(0,0)[cc]{green eyeglass}}
\put(7.33,-0.33){\makebox(0,0)[cc]{red eyeclass}}
\put(54.00,0.00){\makebox(0,0)[cc]{urn}}
\end{picture}
\end{center}
\caption{Wright's generalized urn model put to practical cryptographic use.
\label{f-gum-w} }
\end{figure}
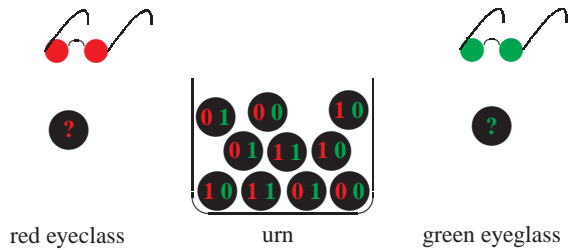

\item[(2)]
After writing down the symbol, Alice hands the chocolate to the quantum, who
in turn carries it to the recipient Bob. During this process, it could,
however, get lost and for some reason never reach its destination (those
with a sweet tooth might for example not be able to wait and eat their
chocolate immediately).

\item[(3)]
Before Bob may take the chocolate and look at it, he, too, needs to flip a
coin in order to choose one pair of glasses. Again, heads is for the green,
and tails for the red ones. He puts them on and takes a look at the
chocolate ball he has just received. He, too, will only be able to read one
of the symbols, as the other one is imprinted in the complementary color and
appears black to him. Then he makes a note of the symbol he has read,
as well as of the color used. As
before, should he attempt to take off his glasses or look at the symbols
with the other pair, the quantum is required to eat the chocolate at once.

After the legal transmission has taken place,
the ``quantum'' may eat the chocolate ball just transfered from Alice to Bob,
or give it away, anyway.

\item[(4)]
Now Bob uses one of the two flags (red or green) to tell Alice whether he
has received anything at all, and what color his glasses are. He does not,
however, communicate the symbol itself.

At the same time, Alice uses one of her flags to inform Bob of the color of her glasses.
Again, she does not tell Bob the symbol she identified.

\item[(5)]
Alice and Bob only keep the symbol if they both received the corresponding
chocolate, and if the color of their glasses (i.e. their flags) matched.
Otherwise, they dismiss the entry.
\end{itemize}

The whole process (1-5) is then repeated several times.

As a result, Alice and Bob obtain an identical random sequence of the symbols 0
and 1. They compare some of the symbols directly to make sure that there has
been no attack by an eavesdropper. The random key can be used in many
cryptographic applications, for instance as one-time pad (like TANs in
online banking).
A more amusing application is to let Alice communicate to Bob secretly
whether (1) or not (0)
she would consider giving him her mobile phone number.
For this task merely a single bit of the sequence they have created is required.
Alice forms the sum $i\oplus j= i+j \,{\rm mod}\, 2$ of her decision and the secret bit
and cries it out loudly over to Bob. Bob can decode Alice's message to plain text
by simply forming the sum $s\oplus t$ of Alice's encrypted message $s$
and the secret bit $t=j$ shared with Alice, for $j\oplus t=0$.
Indeed, this seems to be a very romantic
and easily communicable way of employing one-time pads generated
by quantum cryptography.
(And seems not too far away from the phantasies of its original inventors ;-)

\section{Alternative protocol versions}

There exist numerous possible variants of the dramatization of the BB84 protocol.
A great simplification can be the total abandonment of the black background of the chocolate balls,
as well as the colored eyeglasses. In this case,
both Alice and Bob simply decide by themselves which color to take, and record the
symbols in the color cosen.

In the following, we will present yet another BB84-type protocol
with the context translation principle \cite{svozil-2003-garda}.
First of all, we define one of two possible contexts (either red or green).
Then we randomly measure another context, which is independent of this
choice. If the two contexts do not match (red-green or green-red), a context
translation  \cite{svozil-2003-garda}
is carried out by flipping a coin. In this case, there is
no correlation between the two symbols. If, however, the two contexts match
(red-red or green-green), the results, i.e. the symbols, are identical.

In this protocol, we use sets of two chocolate figures shaped like 0 and 1,
and uniformly colored in red and green, as shown in Table \ref{2005-nl1-t1a}.
An equal
amount of each type of figures is placed inside an urn. No colored glasses
are necessary to carry out this protocol.
\begin{table}
\begin{tabular}{ccc}
\hline\hline
Balltyp&{\Red red}&{\Green green}\\
\hline
Typ 1& --- &{\Green 0}\\
Typ 2& --- &{\Green 1}\\
Typ 3&{\Red 0}& --- \\
Typ 4&{\Red 1}& --- \\
\hline\hline
\end{tabular}
\caption{Coloring and geometry of the four chocolate figures.\label{2005-nl1-t1a}}
\end{table}

The protocol is to be carried out as follows:

\begin{itemize}
\item[(1)]
First of all, Alice randomly draws one figure from the urn and makes a note
of its value (0 or 1) and of its color. Then she gives the figure to one of
the quanta.

\item[(2)]
The quantum carries the figure to Bob.

\item[(3)]
Bob flips a coin and thus chooses one of two colors. Heads is for green,
tails for red. If the color corresponds to that of the figure drawn by Alice
and presented by the quantum (red-red or green-green), the symbol of the
figure counts. If it does not correspond (red-green or green-red), Bob takes
the result of the coin he has just flipped and assigns heads to 0 and tails
to 1. If he wants to, he may flip it again and use the new result instead.
In any case Bob writes down the resulting symbol.

\item[(4-5)]
The rest corresponds to the protocol presented previously.
\end{itemize}

\section{Further dramaturgical aspects, attacks and realization}

It is possible to scramble the protocol in its simplest form and thus the encryption by drawing
two or more chocolate balls, with or without identical symbols on them,
from the urn at once; or by breaking the time order of events.

It is allowed to carry out peaceful attacks in order to to eavesdrop on the
encrypted messages. In the case of the first protocol, every potential
attacker needs to wear colored glasses herself. Note that no one (not even
the quanta) may take additional chocolates or chocolate figures from the
urn, which are identical to the one originally drawn by Alice.
In a sense, this rule implements the no-cloning theorem stating that
it is not possible to copy an arbitrary quantum if it is in a coherent superposition of
the two classical states.

The most promising eavesdropping strategy is the so-called man-in-the-middle
attack, which is often used in GSM networks. The attacker manages to
impersonate Bob when communicating with Alice and vice versa. What basically
happens is that two different quantum cryptographic protocols are connected
in series, or carried out independently from each other. Quantum cryptography
is not immune to this kind of attack.

The first performance of the quantum drama sketched above
took place in Vienna at the  University of Technology as a parallel part of
an event called ``Lange Nacht der Forschung'' (``long night of science'').
Figure \ref{2005-ln1e-pics}a) depicts  the ``quantum channel,'' a ``catwalk''
constructed from yellow painted form liners lifted on the sides with planed wood planks,
through which the individual ``quanta''
had to pass from Alice to Bob.
In the middle of the catwalk, the path forked into two passes, which joined again --
some allusion to quantum interference.
The photographs in Figure \ref{2005-ln1e-pics}b)-f) depict some stages of the performance.

Experience showed that a considerable fraction of the audience obtained some
understanding of the protocol; in particular the players acting as Alice and Bob.
Most people from the audience got the feeling that quantum cryptography is not
so cryptic after all, if they are capable of performing the protocol
and even have fun experiencing it.

For the student of physics probably the most important questions
are those related to the differences and similarities between chocolate balls
and quanta.
This quasi-classical analogy may serve as a good motivation and starting point
to consider the type of complementarity encountered in quantum physics,
and the type of experience presented by single-quantum experiments.

\section*{Acknowledments}
The idea was born over a coffee conversation with G\"unther Krenn.
The first public performance was sponsored by ``Lange Nacht der Forschung''
{\tt http://www.langenachtderforschung.at}.
The chocoate balls ``Mozartkugeln''
were donated by {\it Manner} {\tt http://www.manner.com}.
The black foils covering the balls were donated by
{\it Constantia Packaging}
{\tt http://www.constantia-packaging.com}.
Thanks go to Karin Peter and the public relation office of the TU Vienna for providing the infrastructure,
to the {\it Impro Theater} for the stage performance,
and to Martin Puntigam moderating part of the performances.


\newpage
\begin{widetext}
\begin{center}
\begin{tabular}{lr}
 \includegraphics[width=7.2cm]{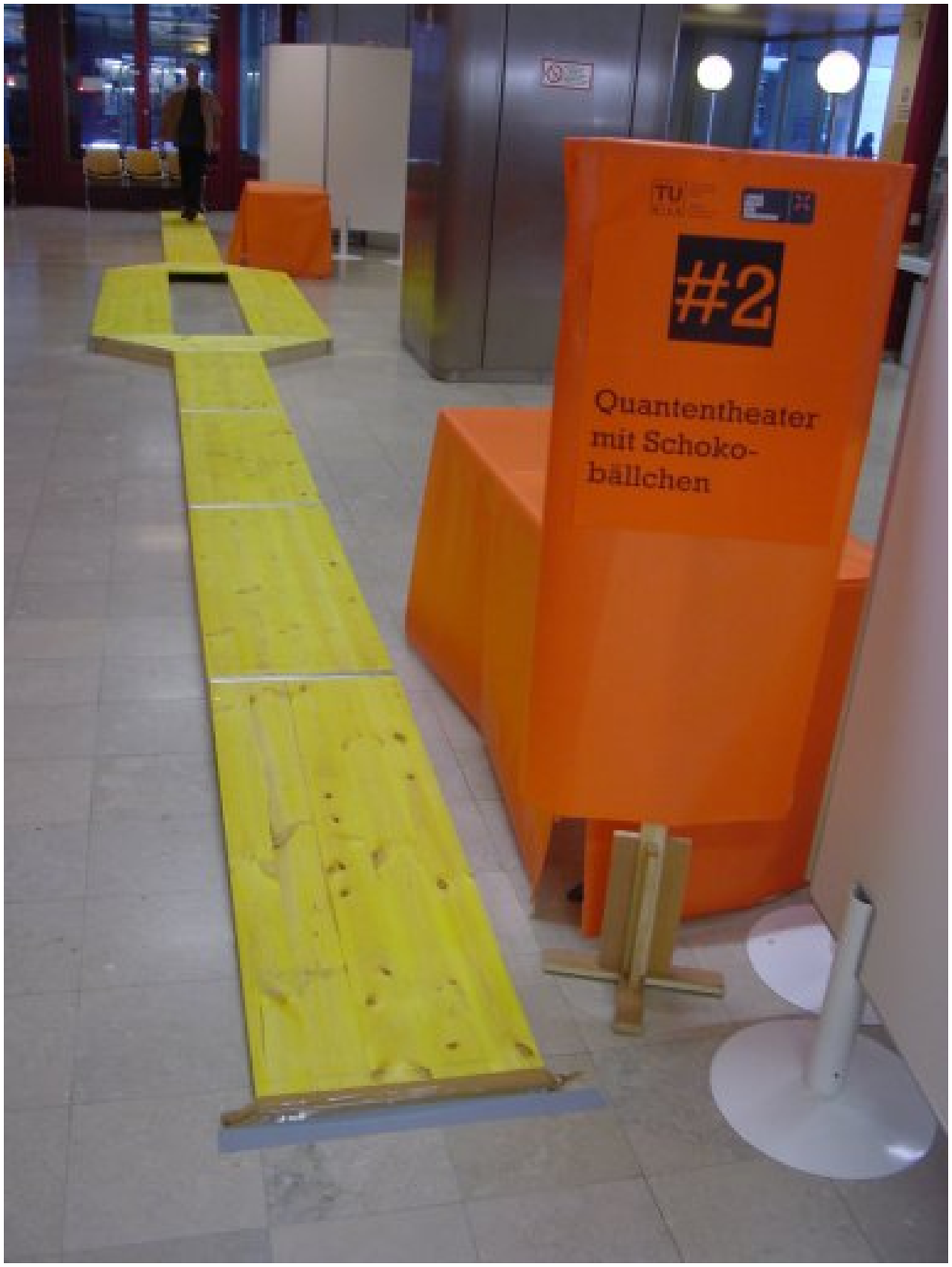}
&
 \includegraphics[width=7.2cm]{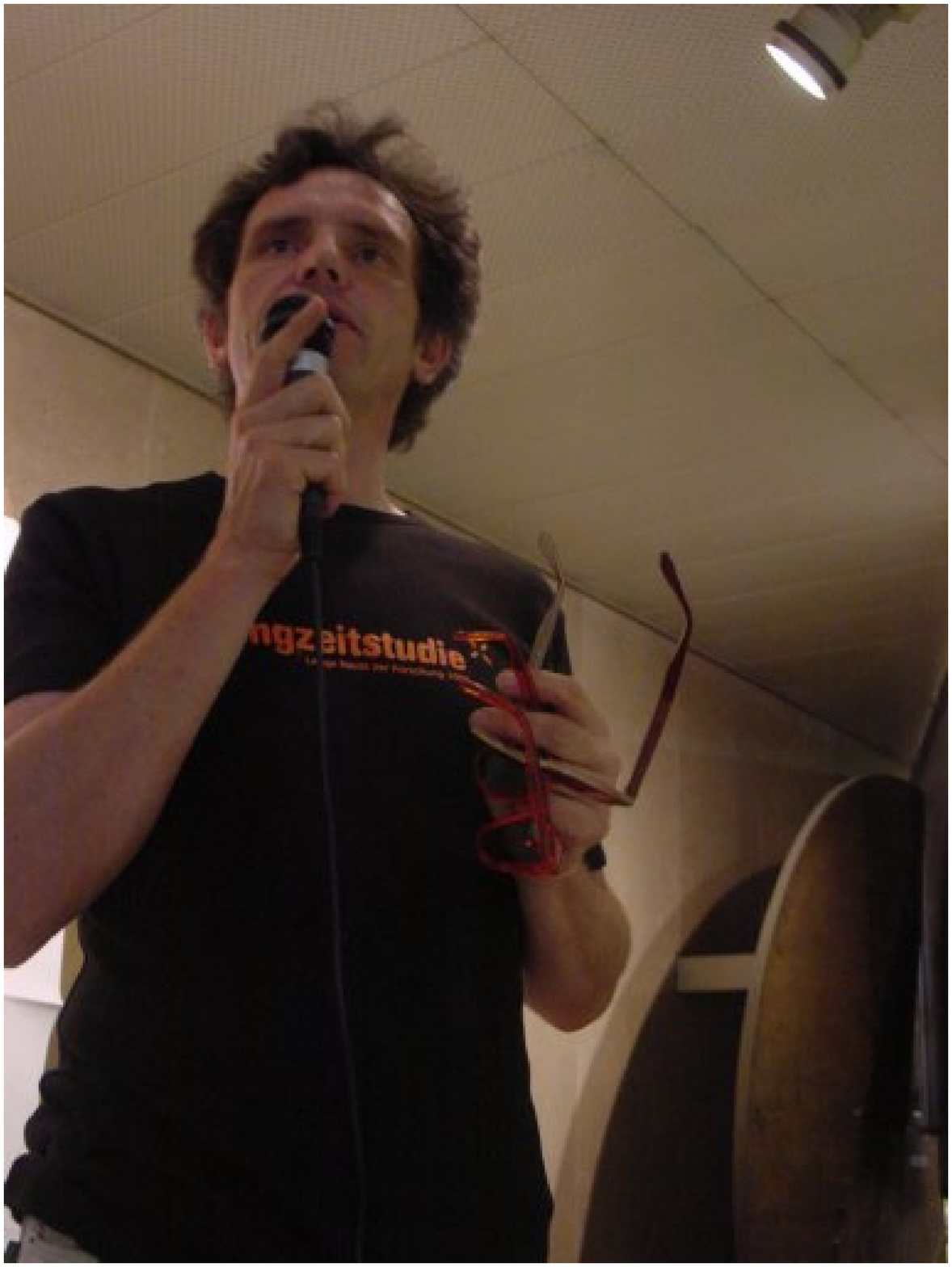}
\\
\multicolumn{1}{c}{(a)}&\multicolumn{1}{c}{(b)} \\
$\;$\\
\end{tabular}
\end{center}
\begin{center}
\begin{tabular}{lr}
\multicolumn{2}{c}{ \includegraphics[width=15cm]{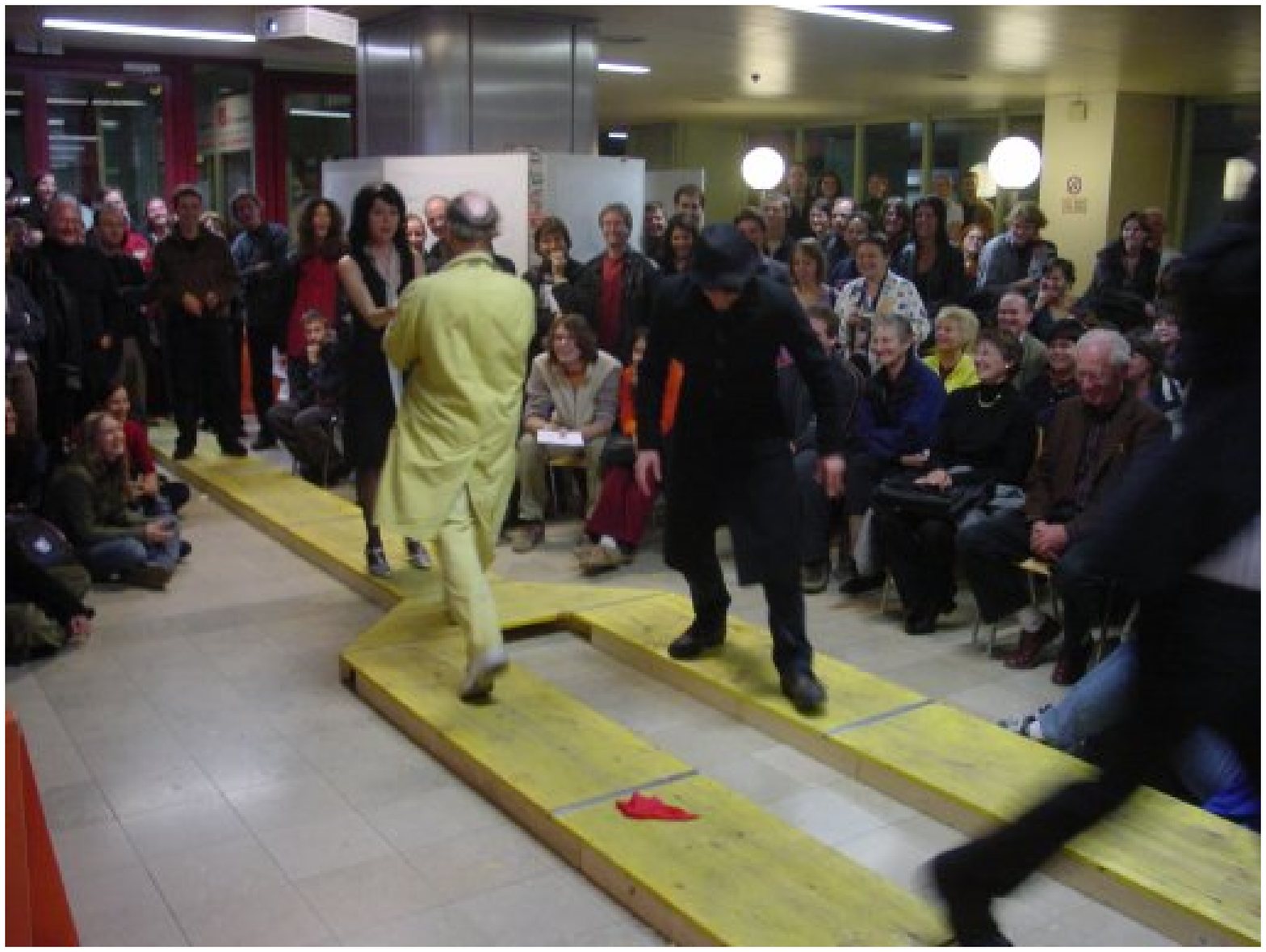}}
\\
\multicolumn{2}{c}{(c)}\\
\multicolumn{2}{c}{ \includegraphics[width=15cm]{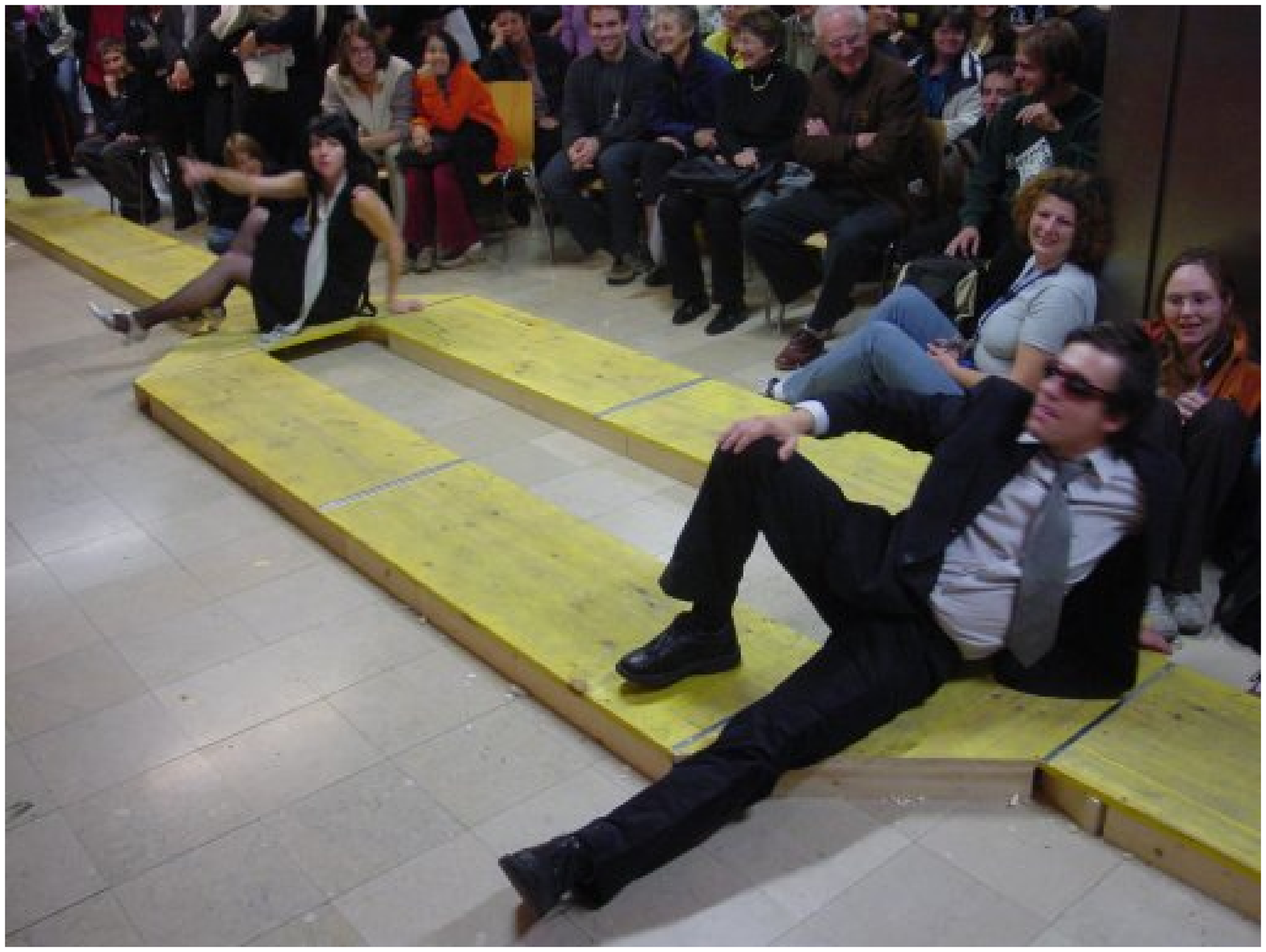}}
\\
\multicolumn{2}{c}{(d)}\\
\end{tabular}
\end{center}
\clearpage
\begin{figure}[t]
\begin{center}
\begin{tabular}{lr}
$\;$\\
\multicolumn{2}{c}{ \includegraphics[width=15cm]{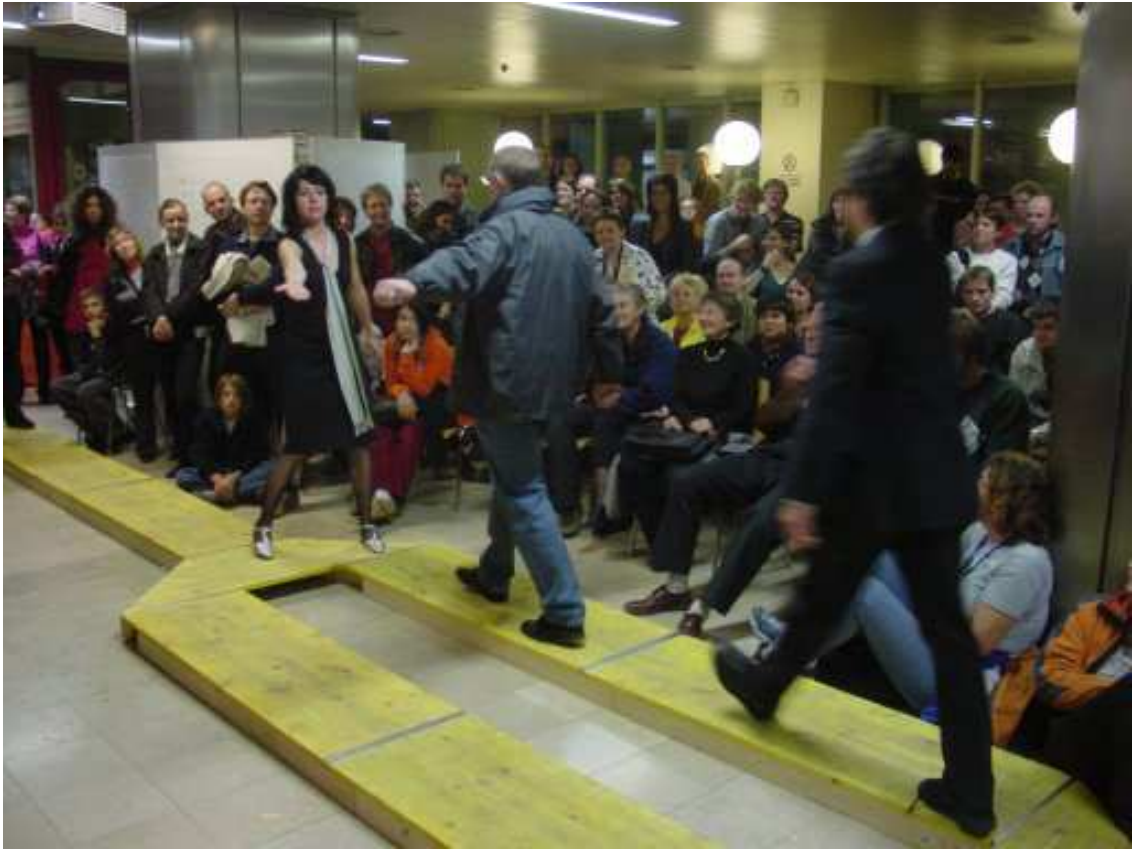}}
\\
\multicolumn{2}{c}{(e)}\\
\multicolumn{2}{c}{ \includegraphics[width=15cm]{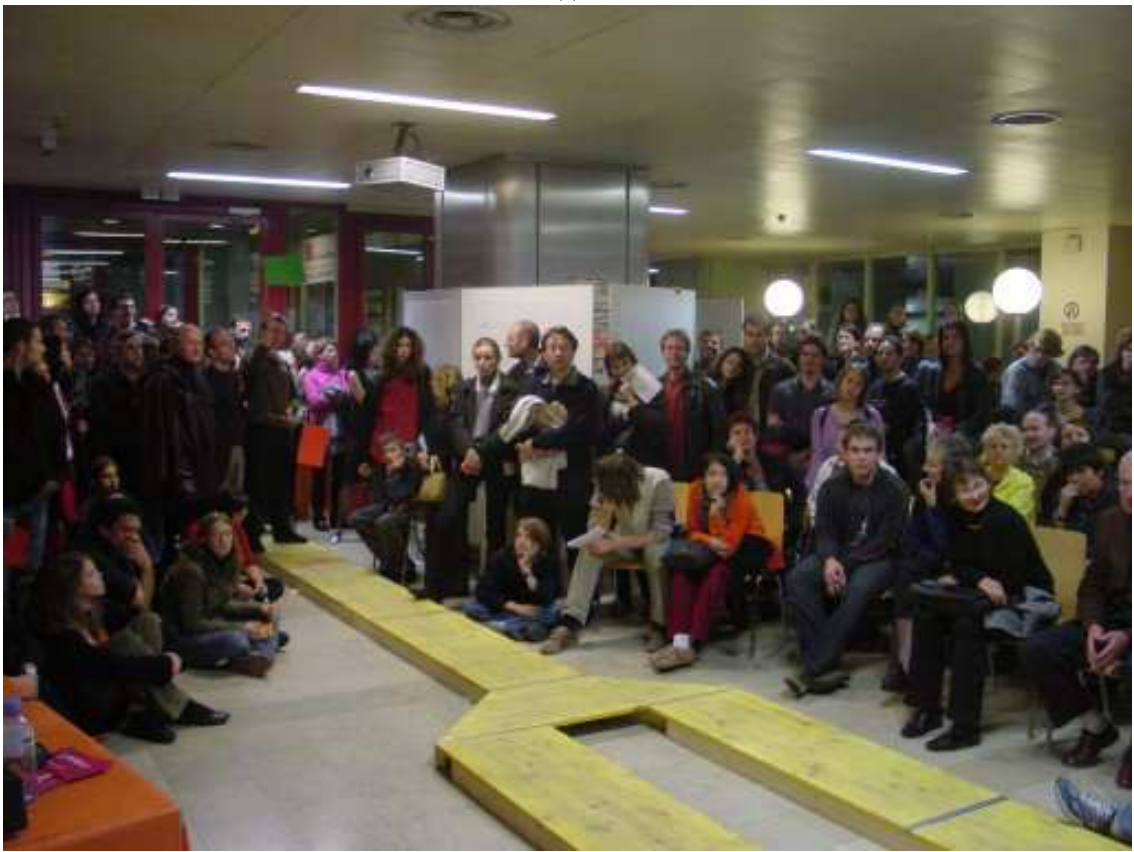}}
\\
\multicolumn{2}{c}{(f)}\\
\end{tabular}
\end{center}
   \caption{Pictures of
(a) the catwalk;
(b) the author with two eyeglasses, one green and one red;
(c) a ``quant'' crossing the catwalk;
(d) agent obstacles in the ``quantum'' catwalk;
(e) another ``quant'' crossing the catwalk;
(f) hissing the flag.
}
\label{2005-ln1e-pics}
 \end{figure}
\end{widetext}
\end{document}